# Molecular Dynamic Approach of Enhanced Self-Propelled Nano-Droplet Motion on Wettability Gradient Surfaces

*Monojit Chakraborty, Anamika Chowdhury, Richa Bhusan, Sunando DasGupta\**

*Department of Chemical Engineering, Indian Institute of Technology, Kharagpur 721302*

**ABSTRACT:** Droplet motion over a surface with wettability gradient has been simulated using molecular dynamics (MD) simulation to highlight the underlying physics. GROMACS and Visual Molecular Dynamics (VMD) were used for simulation and intermittent visualization of the droplet configuration respectively. The simulations mimic experiments in a comprehensive manner wherein micro-sized droplets are propelled by surface wettability gradient against a number of retarding forces. The liquid-wall Lennard-Jones interaction parameter and the substrate temperature were varied to explore their effects on the three-phase contact line friction coefficient. The contact line friction was observed to be a strong function of temperature at atomistic scales, confirming the experimentally observed inverse functionality between the coefficient of contact line friction and increase in temperatures. These MD simulation results were successfully compared with the results from a model for self-propelled droplet motion on gradient surfaces.

## 1. Introduction

Self-propelled droplet motion on gradient surfaces for specialized applications in microfluidic devices is an area of active research in the field of microfluidics and nano-fluidics. Surfaces with wettability gradient have the potential to transport droplets; eliminating the need for pumps and other external devices[1]. Apart from the broad domain of microfluidic devices, mobile droplets are relevant in many new and emerging fields.

Naturally occurring wettability gradient surfaces are found in living organisms in low-rainfall areas to collect water from the humid atmosphere. The Namib desert Stenocara beetle has a patterned wettability surface on its dorsal side[2]. It has alternate hydrophilic bumps and superhydrophobic channels. Water from the atmosphere condenses and nucleates on its hydrophilic bumps and is transported to the mouth of the beetle via the superhydrophobic channels. Such structures can be reproduced and may find application in water-collecting tent, building coverings, water condensers etc. Commercially available superhydrophobic surfacesare being extensively used in the textile industry to develop self-cleaning and water repellent surfaces[3]. The physics of droplet motion on such superhydrophobic gradient surfaces is complex yet worth exploring as the resulting physical insights may lead to new, specialized surfaces.

The applications of gradient surfaces are of diverse nature ranging from biomimetics to heat transfer. Surface energy gradient surfaces are used to investigate the behavior and kinetics of protein adsorption/desorption and relevant in cell attachment, growth and proliferation[4]. Biomimetic studies use these gradient surfaces to develop and test biocompatibility of implants and artificial organs. In condensation heat transfer processes, continuous movement of condensed water droplets away from the cold surface is preferred to maintain high heat transfer coefficient. Wettability gradient surfaces can achieve this with a simultaneous increase in the exposed heat transfer area[5]. They are extremely useful in case of horizontal heat exchange surfaces as well as in microgravity situations. Wettability patterned devices are also used to accurately position functional micro and nano-materials on 2D substrates[3]. The substrate are so patterned that the carrier phase displays distinct wetting/dewetting behavior at specific locations, thereby promoting nucleation at specific, desired locations.

Chaudhary & Whitesides were the first to report experimental observation and measurement of droplet motion against gravity on wettability gradient surfaces[6]. The gravitational force was counter-balanced by the force due to surface tension difference between the two edges of the droplet. The wettability gradient was created by diffusion, deposition and subsequent reaction of DTS vapor on exposed silicon wafer as described by Elwing and Golander[7]. The droplet (of volume of 1-2 micro liters) moved on a uphill plane tilted by $15^o$ at a velocity of ~ 1 to 2 mm/s. Chaudhary and Daniel also demonstrated that droplet movement over wettability gradient surface may be augmented by subjecting it to mechanical vibrations in the form of square waves[8].Droplet motion can also be initiated by the application of thermal gradient. Brzoska et. al[9]. showed that droplet movement was feasible on hydrophobic surfaces by applying thermal gradients. However this movement was reported to occur only above a certain droplet radius known as the critical radius.

Similarly, there have been several other studies for actuating droplet motion using optical[10], acoustic[11] or electrical energy[12].Ichimura et.al. reported induced droplet motion using optical energy[10]. Fluid motion was controlled by photo-irradiation of photo-responsive substrate. The chemical structures of the outermost monomolecular layers were altered by light, which created the wettability gradient. Franke et al. reported motion of droplets in PDMS microfluidic channels using surface acoustic wave device[11]. Rayleigh waves were used to couple with the fluid flow to direct the droplet in the direction of wave propagation. Electrical energy was employed to manipulate droplet motion by Pollack et al.[12] They used two sets of opposing planar electrodes to establish control over the formation and position of micro and nano-droplets.

Droplet motion on a surface with morphological gradient was experimentally studied by Yasuda et. al.[13]. They used lithographic techniques to create patterns of hydrophobic $SiO_2$ and hydrophilic amorphous fluoro polymer which served as the wettability gradient. They reported that steeper gradient in wettability produced larger velocities. Suda & Yamada[14] measured the driving force required to induce droplet movement over gradient surfaces using a flexible glass micro-needle and concluded that the hydrodynamic force was not sufficient to balance the driving force. Instead, they proposed an alternative friction theory wherein the liquid near the substrate may be considered as an assembly of N small solid-like domains. The friction between these domains is said to counteract the driving force.

Brochard[15] theoretically demonstrated that droplet motion can be induced on surfaces with wettability gradient. The gradient can be created by both thermal and chemical means since surface tension ($\gamma$) is a function of temperature as well. A theoretical expression for the velocity of the droplet was also proposed. It was postulated that droplet motion due to the chemical gradient was a result of the drift induced by the solid alone while thermal gradient induced movement was said to be a superimposition of Marangoni flow over the drift induced by the solid. Droplet motion on substrate by the application of thermal gradient was also theoretically reported by Ford & Nadim[16]. They derived the droplet velocity profile induced due to the thermal gradient for drops with small Peclet numbers.

Halverson et. al.[17] used molecular dynamic (MD) simulations to investigate droplet motion on surfaces with wettability gradient. It was observed that the droplets moved steadily for nearly 10 times its base radius for uniform gradient whereas in case of non-uniform gradient droplets were observed to be pinned. The droplet velocities as obtained from simulations were in fair agreement with those predicted from the theory. Inclusion of contact angle hysteresis led to better agreement with the theoretical predictions.

In the present study, we outline the dynamics of nanodroplet movement over gradient surfaces from first principles. A monotonic increase in droplet (composed of ~12000 molecules) motion with temperature is observed. The liquid-wall Lennard-Jones interaction parameter, $\varepsilon_{ls}$, and the substrate temperature have been varied to explore their effect on the three-phase contact line friction coefficient $\zeta$. Changes in $\varepsilon_{ls}$ and substrate temperature cause significant changes in the trajectory of the molecules (expressed through a change in the net molecular displacement frequency, $\kappa^o$) and is used to quantify the variation in $\zeta$. These MD simulation results are successfully compared with the results from a model for self-propelled droplet motion on gradient surfaces.

## 2. Theory

Molecular kinetic theory examines the dynamics of molecules near the three-phase contact line. The driving force for the contact line motion is due to the imbalance of the surface tension forces on wettability gradient surfaces, which disturbs the adsorption equilibrium. The dominant resistive force i.e., three-phase contact line frictional force is molecular in nature since it arises due to the pinning of the molecules at the three-phase contact line. The origin of this resistive force is due to the dissipation of the internal energy of the molecules near the contact line. The liquid molecules adsorbed on the solid substrate, along the three-phase contact line, jump in the forward and backward directions with frequencies $\kappa^+$ and $\kappa^-$ respectively. Velocity of the contact line is given by the net displacement frequency in forward direction times $\lambda$, where $\lambda$ is the average length of one molecular displacement and net displacement frequency is the resultant of $\kappa^+$ and $\kappa^-$[18]. At equilibrium $\kappa^+$ and $\kappa^-$ both are equal and taken as $\kappa^o$. In case of moving wetting line work must be done to overcome the energy barrier of the molecular movement in a specific direction. Driving force for the molecular displacement comes from the out of balance surface tension force $F_w=\gamma(\cos\theta_S - \cos\theta_D)$ at non-equilibrated contact line. Augmenting these ideas gives rise to the expression of forward velocity of contact line as[18,19]

$$U = 2\kappa^0 \lambda \sinh\left[\frac{\gamma(\cos\theta_S - \cos\theta_D)\lambda^2}{2k_B T}\right] \quad (1)$$

Where $\theta_S$, $\theta_D$ are the static and dynamic contact angles, $\gamma$ is the liquid-air interfacial tension, $\kappa^o$ is the equilibrium frequency of random molecular displacements and is a function of activation free energy of wetting[18], $\lambda$ is the average length of one molecular displacement. If the argument of hyperbolic sine is very small, the equation reduces to

$$U = \frac{\kappa^0 \gamma(\cos\theta_S - \cos\theta_D)\lambda^3}{k_B T} \quad (2)$$

Where $\kappa^o$, $\gamma$ and $\lambda$ are clubbed together in a single parameter $\zeta$ known as the three-phase contact line friction coefficient, defined as[19],

$$\zeta = \frac{k_B T}{\kappa^0 \lambda^3} \quad (3)$$

It has been extensively reported in the literature that $\zeta$ can only be determined using experimental data. It is clear from equation (3) that $\zeta$ depends on temperature and $\kappa^o$. Molecular dynamics simulation is used to probe the effect of these parameters on $\zeta$ and on the dynamics of droplet motion on gradient surfaces.

## 3. Simulation Details

The simulations reported herein were carried out using GROMACS-4.0.7[20-21-22] whereas Visual Molecular Dynamics (VMD)[23] was used for intermittent visualization of the droplet configuration. The well-known L-J [$U_{LJ}(r)$] potential was used to model the system,

$$U_{LJ}(r) = 4\varepsilon_{ij}\left[\left(\frac{r}{\sigma}\right)^{-12} - \left(\frac{r}{\sigma}\right)^{-6}\right] + \sum_{i=1}^{n}\sum_{j=1}^{n}\frac{k_e q_i q_j}{r_{ij}} \quad (4)$$

Where $\varepsilon_{ij}$= depth of the potential well; $\sigma$= finite distance at which inter-atomic potential is zero; r=inter-atomic distance; $k_e$=electrostatic constant; $q_i$, $q_j$= charge on $i^{th}$ and $j^{th}$ atom; $r_{ij}$ = distance between $i^{th}$ and $j^{th}$ atom;

Simple point charge (SPC) model[24] has been employed for the simulations, where in the water molecules assume ideal tetrahedral shape with a H-O-H bond angle of $109.47^o$. Oxygen molecules and the oxygen-wall interactions were assumed to be of L-J type. The interactions between the solid-solid atoms were excluded. For the SPC water model, $\sigma_{oo}$=0.3166 nm[25] and for Si atoms, $\sigma_{si}$ =0.3408 nm[26].

According to the Lorentz-Brethelot mixing rule[27]

$$\sigma_{ls} = \left(\frac{\sigma_s + \sigma_l}{2}\right) \quad (5)$$

Thereby for the reported simulations, the value of $\sigma_{ls}$ has been fixed to 0.3276 nm and as a thumb rule a cut-off length of 1.2 nm ($3.66\sigma_{ls}$)[27] has been used for coulombic and van der Waals forces. Periodic boundary conditions were applied in both X and Y directions. The temperature of the system was maintained using v-rescale coupling algorithm[28] with a time-constant of 0.1ps for the water droplet. The force field applied was 53a6 of GROMOS-96 version[29]. The integrator used was steepest descent type. Simulations for the movement of water molecules on gradient surfaces were performed for 3000ps and the spreading dynamics for the same was observed for 2000ps. The dimensions of the simulation box used were 35x35x10 (nm) and comprising of around 12000 water molecules was found to be optimum based on computation time and simulation results. The simulations were performed using NVT (at different constant temperatures) ensemble[30,31].

*3.1 Spatial variation of Epsilon-* The magnitude of the interaction parameter ($\varepsilon$) is an indicator of the attractive force between the atoms and its strength is directly proportional to the substrate hydrophilicity. To study the variation of contact angle with epsilon, water drop equilibration on homogenous surfaces was simulated for different epsilon ($\varepsilon_{ls}$) values. The equilibrium contact angles were calculated from the extracted images using ImageJ software. The contact radius was computed using the equilibrated drop coordinates obtained from the simulation. The base layer molecules were identified from their z-coordinates and the contact radius was computed. The variation in contact angle and contact radius as a function of the interaction parameter is shown in Figure 1.

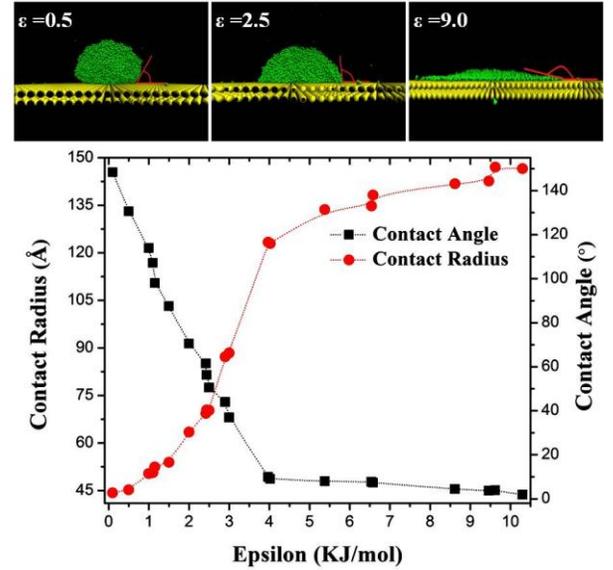

**Figure 1** Contact Angle and Contact Radius variation as a function of Epsilon

Reduction in contact angle with increase in epsilon corroborate to the fact that substrate hydrophobicity decreases with increase in attractive forces between the substrate and water droplet. The other component of the interaction potential ($\sigma$) cannot be altered since the minimum cut-off distance specification imposes a constraint on the inter-particle distance. The interactions among the wall molecules were excluded while performing the simulation studies.

*3.2 Creation of Gradient Surface-* A wall (34x34 (nm)) consisting of two layers of Si atoms in cubic crystal structure in the x-y plane was created. Parallel strips running in the y-direction were used to mimic gradient in surface tension and to function as the gradient surface. The values of the interaction parameter ($\varepsilon$) for these parallel strips are set as shown in Table 1 to create a surface with appropriate gradients in wettability to induce droplet motion. The variation in the difference in interaction parameter (Δε) is manifested not only in the motion of the droplets but will be such so as to mimic the behaviour of the driving force acting on droplets to generate velocity profiles as encountered in experiments[32]. It has been shown that the velocity of these droplets increases rapidly, reaches a maximum and ultimately slows down gradually till the end of the droplet motion[32,33]. The variation in the interaction parameter (ε) is chosen such that the resulting motion is qualitatively similar. The gradient strips were limited in the y-direction to a size of 12 nm to ensure droplet spreading in one (x) direction only. The simulation runs were performed for the water droplet, keeping the wall atoms frozen at their positions.

**Table 1**: ε values as a function of x-coordinate to create a wettability gradient surface

| Wall Coordinates (A°) | ε (KJ/mol) | Δε (KJ/mol) |
|---|---|---|
| 0 – 50 | 0.1 | 1.05 |
| 50 – 80 | 1.15 | 1.35 |
| 80 – 110 | 2.5 | 1.65 |
| 110 – 140 | 4.15 | 1.5 |
| 140 – 170 | 5.65 | 1.3 |
| 170 – 200 | 6.95 | 1.14 |
| 200 – 230 | 8.09 | 1.04 |
| 230 – 260 | 9.13 | 0.96 |
| 260 – 290 | 10.09 | 0.92 |
| 290 – 340 | 11.01 | |

## 4. Results and Discussion

Droplet motion has been implemented over the created gradient surface with a run time of 3000ps for each of the simulations.

### 4.1 Droplet Motion Simulation on Gradient Substrate

The cubical block of water molecules was initially positioned towards the hydrophobic end of the gradient wall. The temperature was kept constant at 300K. Figure 3 represents the droplet motion from the hydrophobic end towards the hydrophilic end on the substrate. It can be observed that the mean density of the simulated droplet remained close to the actual density of water (see supporting information for details) throughout the simulation run

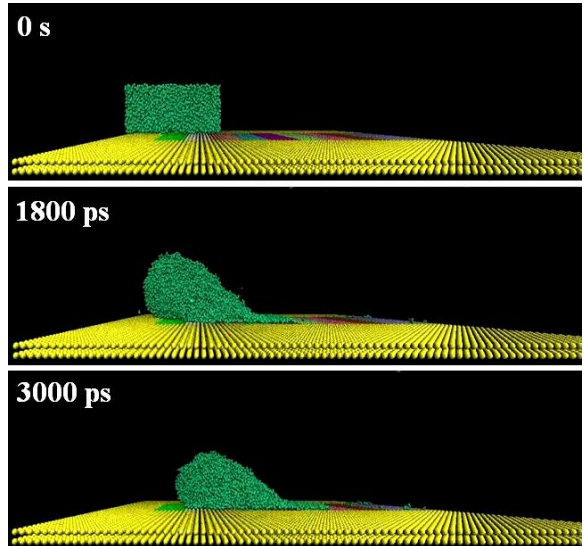

**Figure 3** Simulation of droplet motion at 300K.

*4.2. Effect of interaction parameter (ε) on velocity profile of the droplet:*

Three different substrates of varying hydrophilicity gradients (Δε/Δx) were adopted for each simulation. The variation of the gradient of Δε with mean coordinate of each strip is shown in Figure 4. The substrate designated by Test 3 has the stiffest Δε gradient.

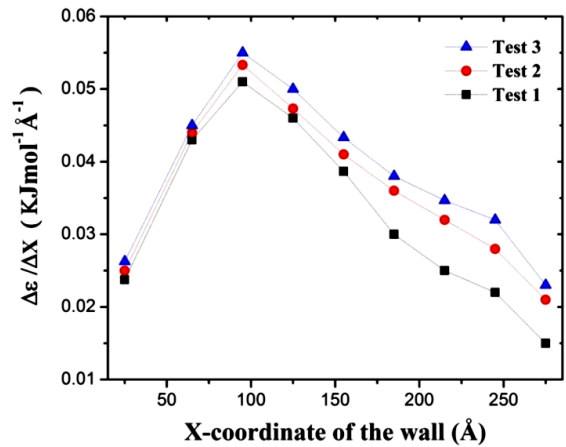

**Figure 4** Variation of Δε gradient for three different test substrates

The co-ordinates of the center of mass along the direction of movement of the droplet were determined at every time instant from the output trajectory file, where the center of mass is defined as that coordinate which minimizes the mass weighted sum of a periodic spatial function such as a cosine function for the group of atoms which are eventually a part of a larger system[34].

The displacements of the center of mass along the direction of wettability gradient were evaluated. It was observed that the displacements along the other directions were negligible(see supporting information).The instantaneous velocities were obtained by point-wise differentiation of the fitted displacement curve along the direction of the motion and are presented in Figure 5.It is apparent that the peak velocities recorded were strongly influenced by the stiffness of the gradient in Δε. It is clear from Figures 4 & 5 that droplet motion can be controlled by tuning the Δε gradient.

The velocity profile, obtained from the simulation results, initially showed an increasing trend till it reached a maximum and then started to decrease. The characteristics and trends obtained from these simulations were remarkably similar to the experimental trends and in accordance with the displacement-time curve inflection point[32] with the driving force being proportional to Δε. The velocities obtained in the simulations are of the order of 2.5 – 3 m/s which are an order of magnitude higher as compared to the velocities reported in experiments[33]. This may be attributed to the scaling down of the droplet size (from mm to nanometer, resulting in a decrease of total mass of the order of~$10^{-16}$) which reduces the inertial resistance. It has been reported before that significant increases in the droplet speeds (up to three orders of magnitude) were encountered on decreasing the size of the droplet to nanometer[17].

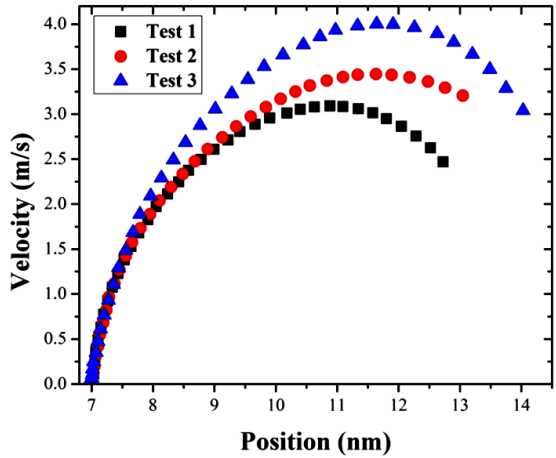

**Figure 5** Velocity Profiles for three different tests at 300K; Simulation data at every 20th point are shown for clarity.

### 4.3 Effect of Temperature Increase:

Simulations at different temperatures (in steps of 2K from 298 K to 310 K) were performed on the gradient surface designated as Test 3(Figures 4 & 5) to study the effect of temperature change on molecular movement. The maximum temperature used was limited to 310 K since higher temperatures would cause significant loss of water molecules due to the evaporation. The simulation results are presented in Figure 6 and Table 2, showing a distinct increasing trend of the droplet movement with temperature.

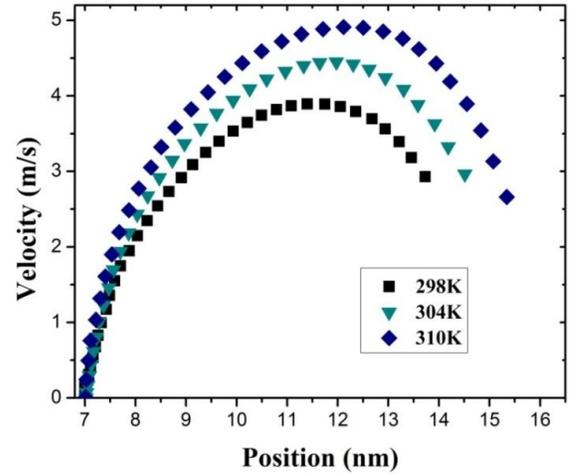

**Figure 6** Velocity variation with temperature for Test 3; Simulation data at every 20th point are shown for clarity.

**Table 2** Variation of peak velocity with temperature for Test 3

| Temperature (K) | Peak Instantaneous Velocity (m/s) |
|---|---|
| 298 | 3.896 |
| 300 | 4.003 |
| 302 | 4.118 |
| 304 | 4.450 |
| 306 | 4.580 |
| 308 | 4.712 |
| 310 | 4.913 |

The increase in velocity with temperature is attributed to the decrease in friction associated with the contact line motion. The contact line friction coefficient is a combination of several parameters and has been estimated in the following section to gauge its response to surface hydrophilicity (which is characterized through the interaction parameter $ε_{ls}$) and temperature.

### 4.4 Effect of surface hydrophilicity and temperature on contact line friction coefficient

The values of the three-phase contact line friction coefficient, ζ were calculated for a series of simulation of droplet motion on a uniform wettability surface at different temperatures. The simulation was also run on surfaces having different uniform wettability (different values of ε) at a constant temperature. The procedure adopted for the evaluation of ζ calculation is similar to that used by Ruijter et.al.[35]. Equation 3 was used to calculate ζ by analyzing the MD simulation results. In terms of molecular displacement, the interface ad-vancement can be represented as molecules jumping from one site to another with a frequency $\kappa^o$ and distance $\lambda$. Near the wall, the molecular jumps were anisotropic in nature, while far away they were equal in all directions. The jump frequencies can be divided into three types[35]; the parallel frequency($\kappa_{\parallel}$) of the molecules of the first layer, the perpendicular frequency ($\kappa_{\perp}$) of the first layer and the bulk frequency ($\kappa_{Bulk}$).

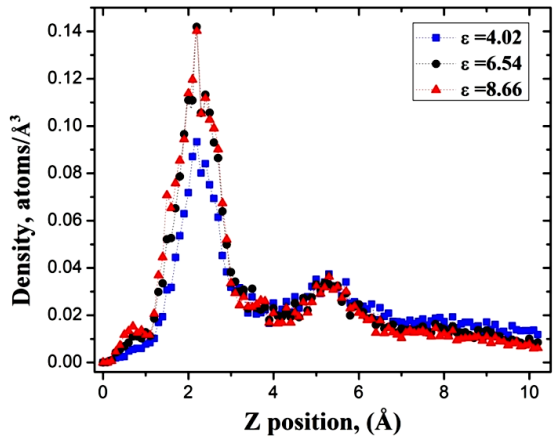

**Figure 7** Density variation with distance from the wall surface plotted at 298K.

To identify the first two layers of the water molecules, the atomic density profile in the direction normal to the wall surface has been plotted[36] (Figure 7), while keeping the temperature constant at 298K. The density profile of the equilibrated drop has been evaluated at 1600 ps as no significant change in the droplet profile was observed except minor spreading. It is clear from the figure that close to the surface of the wall, the droplet showed well defined water layers with distinctly high number density. However, the number density dropped farther from the wall and after about 10Å became equal to the bulk number density. It is clear that with increase in the value of $\varepsilon$ (increasing hydrophilicity) the number density of the first layer increased marginally. However, with increase in temperature (with constant $\varepsilon$) the first layer density remained almost constant.

The first layer of water molecules for all simulation combinations were identified (using Figure 7) and their planar displacements were noted. A parallel jump is defined as the displacement of a water molecule not less than $\lambda_{\parallel}$ in planes parallel to the wall surface. The jump of the molecules near the wall is attributed to the adsorption phenomenon taking place on the wall surface[37]. Thus $\lambda_{\parallel}$ was approximated as the largest distance between two consecutive wall atoms (adsorption sites)[37], which was $4\sqrt{2}$Å. The number of molecules at a particular time instant which achieved a parallel jump for the first time was noted and the cumulative percentage of such molecules was plotted against time (Figure 8) at different values of surface hydrophilicity ($\varepsilon$) at a constant temperature at 298K. Figure 9, on the other hand, is a similar plot with varying temperature at constant surface hydrophilicity ($\varepsilon$= 6.54; contact angle 7.67°)

The cumulative time required for half of the molecules to get displaced by at least $\lambda_{\parallel}$ was noted (dashed line in Figure 8 and Figure 9). The value of $t_{\parallel}$ was taken to be half of this time to take into account the possibility of both forward and backward displacements[37] and $\kappa_{\parallel}$ was simply the inverse of $t_{\parallel}$.

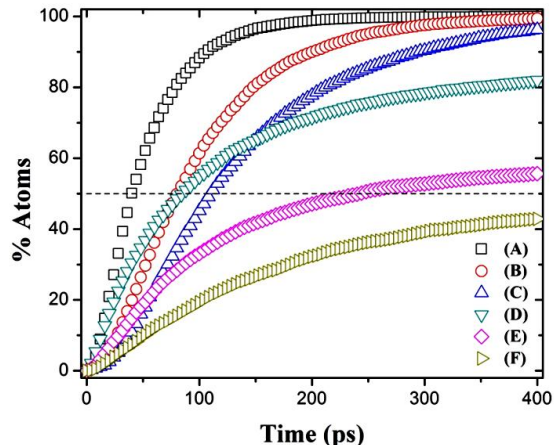

**Figure 8** Cumulative percentage of water molecules displaced by $\lambda$ as a function of time at 298K, while varying interaction parameter $\varepsilon$. (A), (B), (C) are cumulative percentage parallel displacement for $\varepsilon$ = 4.02, 6.54 and 8.66 respectively and (D), (E) and (F) are cumulative percentage perpendicular displacement for $\varepsilon$ = 4.02, 6.54 and 8.66. Only three values of $\varepsilon$ are shown, to enhance readability.

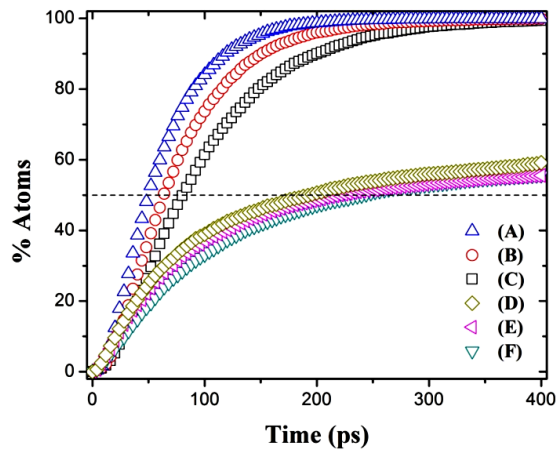

**Figure 9** Cumulative percentage of water molecules displaced by $\lambda$ as a function of time while varying temperature at constant surface hydrophilicity ($\varepsilon$=6.54; Contact angle 7.67°). (A), (B), (C) are cumulative percentage parallel displacement at 310 K, 304 K and 298 K respectively and (D), (E) and (F) are cumulative percentage perpendicular displacement at 310 K, 304 K and 298 K Only three values of temperature are shown, to enhance readability

The perpendicular frequency ($\kappa_\perp$) was computed for all the atoms jumping from the first to the second layer. The distance between the first two crests in the number density plot (from Figure 7) was noted as $\lambda_\perp$[37]. A perpendicular jump is defined as the displacement greater than $\lambda_\perp$ in the direction perpendicular to the wall plane[37]. The cumulative percentage graph (Figures 8&9) was plotted using similar approach as adopted for the evaluation of $\kappa_\parallel$. The perpendicular jump frequency, $\kappa_\perp$ was calculated by inversing the time required for 50% of the molecules to get displaced by at least $\lambda_\perp$.

A sharp initial increase in cumulative percentage was noted after which the curve tapered off and reached saturation. It was observed that with increase in ε, the percentage of molecules displaced at a particular instant of time decreased. Increases in ε amplified the attractive forces between the water molecules and the substrate thereby resisting displacement of molecules. The molecules require more energy to detach from its location and jump to another site, thereby reducing the frequency of molecular displacements with increase in ε.

The percentage of molecules displaced from their original sites increases with increase in temperature resulting in an increase in frequency. Further, it can be seen both in Figure 8&9 that the molecular displacement frequency in the parallel direction is about twice that of the perpendicular direction ($\kappa_\parallel \sim 2\kappa_\perp$) even though the former had a higher value of jump distance. This difference can be accredited to the additional degree of freedom in the x-y plane[37].

The bulk movement of molecules can be characterized by a distance $\lambda$ and a frequency $\kappa^o$. The distance $\lambda$ accounted for displacements in all the three directions, and hence the limiting value of $\lambda_{Bulk}$ can be approximated by the average intermolecular distance of the water molecules.[35] For a movement to be considered as a successful bulk jump, the displacement in any direction must be greater than $\lambda_{Bulk}$. The average intermolecular distance was calculated from the initial cubical arrangement of the water molecules. A total of 12660 water molecules were randomly accommodated in a cubical arrangement of 100 x 100 x 40 (Å). This initial arrangement of water molecules can be assumed to be equivalent to a cuboidal box of same dimensions with molecules arranged in a regularly spaced pattern such that each individual molecule occupies the center position of a cubical volume equivalent to 0.0316 nm³. The intermolecular distance was calculated using the box dimension and was found to be equal to 3.16 Å. This was taken as the limiting value of $\lambda_{Bulk}$. The cumulative percentage of water molecules displaced by a distance $\lambda_{Bulk}$ was plotted as a function of time (Figures 10 & 11) and was used to calculate the average frequency of molecular displacement ($\kappa_{Bulk}$).

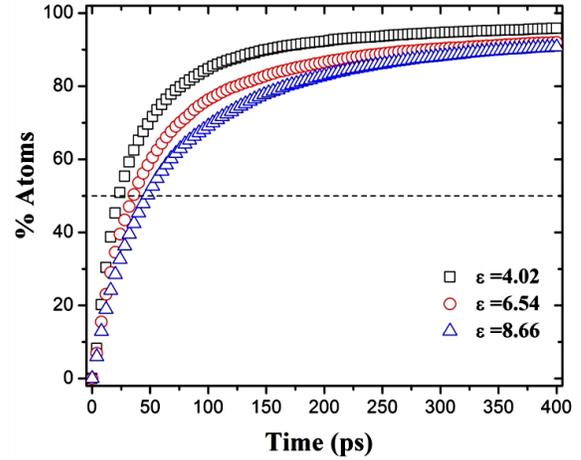

**Figure 10** Cumulative percentage of water bulk movement molecules displaced by unit distance as a function of time while varying interaction parameter ε and at constant temperature 298K, Only three values of ε are shown, to enhance readability

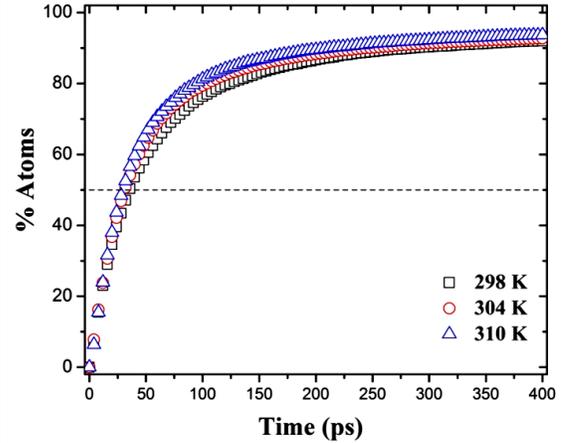

**Figure 11** Cumulative percentage of bulk movement of water molecules displaced by unit distance as a function of time while varying temperature at a constant ε=6.54. Only three temperatures are shown, to enhance readability.

$\kappa_{Bulk}$ is defined as the inverse of the time needed to displace 50% of the molecules by one intermolecular distance from their initial position. The above procedure was repeated with different values of ε and temperature.

The rate obtained for parallel jumps is higher than that of perpendicular and bulk displacements for reasons already described. The contact line movement is governed by the water molecules of the first two layers and, majority of the molecules undergo parallel displacements at any given time. Thus, $\kappa_\parallel$ was used for computing $\zeta$[35-37]. Three-phase contact line friction coefficient, $\zeta$ was computed for different ε and temperature using equation 3 and has been plotted in Figure 12. The obtained patterns depict appreciable reduction in $\zeta$ with temperature. Since $\zeta$ is a measure of the contract line friction, which in turn depends on the attractive force between the droplet and the substrate (more for the hydrophilic substrates, larger $\varepsilon$), significant increase in $\zeta$ with increases in $\varepsilon$ are observed. Even though the defining equation for $\zeta$ (Equation 3) shows a direct relationship with temperature, the amplification in frequency of molecular displacement with temperature more than compensates for the increases of $\zeta$ with temperature resulting in a decreasing trend with temperature.

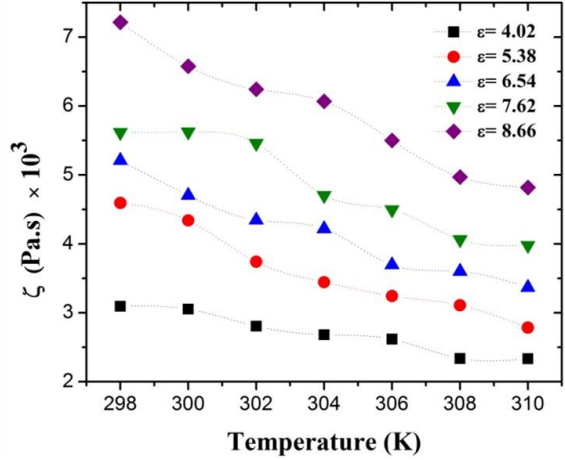

**Figure 12** Variation of $\zeta$ as a function of temperature and $\varepsilon$.

4.5 Comparison with the analytical model

Theoretical predictions of droplet velocities were made based on the approach used by Chakraborty et al.[33] and were compared with the values obtained from the simulation results of the present study. The instantaneous velocity of the droplet on a wettability gradient surface can be evaluated by equating the surface tension gradient inspired driving force and the resistive forces acting on the drop, namely the hydrodynamic and the three phase contact line forces using a quasi steady state approximation[38].

$$F_{driving} = F_h + F_{cl} \quad (6)$$

where, $F_h$ = hydrodynamic force; $F_{cl}$= three-phase contact line force;

The droplet also experiences a drag force if it is encapsulated by a filler medium. However, this drag force was found to be insignificant when compared to the other resistive forces and therefore has been neglected[33]. Brief descriptions of these forces are given below.

*Driving force:* The change in the surface energies is quantified as the difference in the receding and the advancing contact angles at various instants of time at every position and is used to evaluate the driving force at that location. It can be expressed as[39],

$$F_{driving} = 2R(x)\gamma g(\theta) \quad (7)$$

where $g(\theta) = \int_0^{\pi/2} [\cos(\theta)_a - \cos(\theta)_r] \cos\phi \, d\phi$

Where, R(x) is the radius of the footprint of the drop and $\gamma$ is the liquid-gas interfacial tension.

*Hydrodynamic force:* Subramanian et al.[39] used a wedge approximation to the Stokes flow condition analysis by Cox[40] to predict the hydrodynamic force exerted by the solid surface (silicon wafer) on the liquid drop as,

$$F_h = (8\mu U(x) R(x) f(\theta, \varepsilon)) \quad (8)$$

where,

$$f(\theta,\varepsilon) = \int_0^{(1-\varepsilon)} \frac{(1-Y^2)\tan^2\theta[\frac{1}{2}\ln((1-Y^2) - \ln\varepsilon]}{[\tan\theta\sqrt{(1-Y^2)} - (1+\{1-Y^2\}\tan^2\theta)\tan^{-1}(\tan\theta\sqrt{(1-Y^2)})]} dY$$

$U(x)$ = the instantaneous velocity of the drop
$\theta$ = the dynamic contact angle of the drop.

and $\varepsilon = \dfrac{L_s}{R}$

where, $L_s$ is the slip length[32]. Molecular dynamics simulations depicting similar physical situations as in the present study indicated slip length values are of the order of 0.5 nm[32].

*Three-phase contact line Friction Force:* As the droplet moves along the path of decreasing hydrophobicity, dynamic wetting can be visualized as the displacement of the three phase contact line from one location to another. The advancement of three phase contact line leads to dissipation of energy at the molecular level[18] expressed by the coefficient of the contact line friction. Thus,

$$F_{cl} = 2P(x)\zeta U(x) \quad (9)$$

Where,
$P(x)$ = the droplet perimeter length
$\zeta$ = the coefficient of contact line friction[41-42]

Substituting equations (7), (8) and (9) in equation (6) a linear equation in $U(x)$ is obtained as,

$$2R(x)\gamma g(\theta) = 8\mu U(x) R(x) f(\theta,\epsilon) + 4\pi R(x)\zeta U(x) \quad (10)$$

leading to the expression of $U(x)$,

$$U(x) = \frac{2R(x)\gamma g(\theta)}{8\mu R(x) f(\theta,\epsilon) + 4\pi R(x)\zeta} \quad (11)$$

For a given solid–liquid system, there is no definitive way of predicting the values of $\zeta$. Thus, it is impossible to predict the dynamic wetting behavior from independently measured quantities[42] for a dynamic system. Hence, the values of the parameter $\zeta$ needed to be evaluated in situ for each simulation. Here, the values of $\zeta$ were evaluated by minimizing the root mean square error between the theoretically predicted velocities (equation 11) with the simulation (Figure 6) results. The value of ζ as reported in the literature vary considerably, from 0.000089 Pa-s for reported molecular dynamics simulations of dynamic wetting behavior of liquid to 738000 Pa-s for melting of metal and oxides on Molybdenum[43,44-45-46].

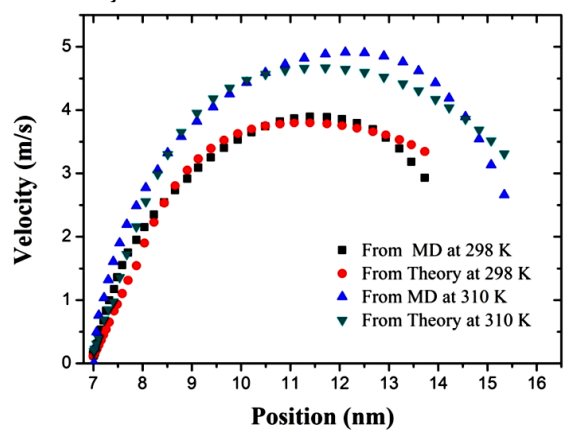

**Figure 13** Comparison of velocities between simulation and theoretical studies (Droplet velocities are calculated using Equation 11). Data at every 20th point are shown for clarity. To enhance readability, comparisons at only two temperatures are presented here.

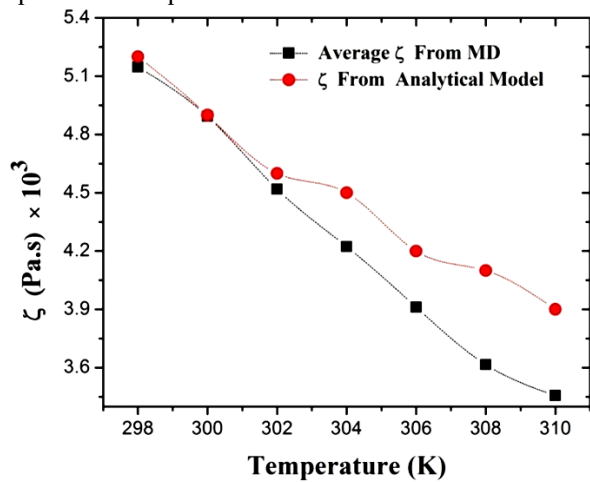

**Figure 14** Comparison of ζ between simulation and theoretical studies

For the simulations reported herein, the value of ζ was found to be in the range of 0.0035- 0.0052 Pa.s, comparable with the orders reported by the researchers[43-46] for similar studies. The velocity profiles and ζ values obtained using equation (11) and those from simulation results (Figures 6 & 12) have been plotted and compared in Figure 13 and 14 with variations in ζ similar to that as observed in the experiments[33]. It is clear that an increase in temperature alters the molecular movements at the three phase contact line friction leading to a decrease in ζ causing faster droplet movement. This may lead to specific applications involving temperature induced enhanced droplet manipulations at the micro/nanoscale.

## 5. Conclusions

Molecular dynamic simulations of wettability gradient induced water droplet motion on a solid substrate are performed using GROMACS and Visual Molecular Dynamics (VMD). The driving force for the molecular displacement is a result of out of balance surface tension forces at the non-equilibrated contact line. The simulations are performed with 12000 water molecules at a time scale of 3000 ps. The gradient surface is created using different values of the interaction parameter to mimic reported experimental trends in droplet motion (with a maximum velocity of the order of 2.5 – 3 m/s). The simulation results demonstrate a distinct increasing trend of the droplet movement with temperature (an increase of ~1 m/s for a 10 K rise in temperature). This increase is attributed to the decrease in contact line friction wherein the relevant parameter ζ decreases from 0.0052 Pa.s to 0.0035 Pa.s. A force balance method is proposed where the wettability gradient induced driving force is balanced by the hydrodynamic and the contact line friction force with the later being the dominant one. The results predicted from the theory are successfully compared to that of the MD simulations including the variations of ζ with temperature.

## AUTHOR INFORMATION

**\*Corresponding Author**

Email:     sunando@che.iitkgp.ernet.in,
Ph: +91 - 3222 - 283922

## REFERENCES

(1) Grunze, M. Driven Liquids. *Science* **1999**, *283*, 41–42.
(2) Parker, a R.; Lawrence, C. R. Water Capture by a Desert Beetle. *Nature* **2001**, *414*, 33–34.
(3) Yao, X.; Song, Y.; Jiang, L. Applications of Bio-Inspired Special Wettable Surfaces. *Adv. Mater.* **2011**, *23*, 719–734.
(4) Choi, S.; Newby, B. Z. Micrometer-Scaled Gradient Surfaces Generated Using Contact Printing of Octadecyltrichlorosilane. *Langmuir* **2003**, *19*, 7427–7435.
(5) Daniel, S.; Chaudhury, M. K.; Chen, J. C. Fast Drop Movements Resulting from the Phase Change on a Gradient Surface. *Science* **2001**, *291*, 633–636.
(6) Chaudhary, M. K.; Whitesides, G. M. How to Make Water Run Uphill. *Science* **1992**, *256*, 1539–1541.
(7) Elwing, H.; Golander, C.-G. Protein and Detergent Interfacial Phenomena on Solid Surfaces with Gradients in Chemical Composition. *Adv. Colloid Interface Sci.* **1990**, *32*, 317–339.
(8) Daniel, S.; Chaudhury, M. K. Rectified Motion of Liquid Drops on Gradient Surfaces Induced by Vibration. *Langmuir* **2002**, *18*, 3404–3407.
(9) Brzoska, J. B.; Brochard-Wyart, F.; Rondelez, F. Motions of Droplets on Hydrophobic Model Surfaces Induced. *Langmuir* **1993**, *9*, 2220–2224.
(10) Ichimura, K. Light-Driven Motion of Liquids on a Photoresponsive Surface. *Science* **2000**, *288*, 1624–1626.

(11) Franke, T.; Abate, A. R.; Weitz, D. A.; Wixforth, A. Surface Acoustic Wave ( SAW ) Directed Droplet Flow in Microfluidics for PDMS Devices. *Lab Chip* **2009**, *9*, 2625–2627.

(12) Pollack, M. G.; Shenderov, a D.; Fair, R. B. Electrowetting-Based Actuation of Droplets for Integrated Microfluidics. *Lab Chip* **2002**, *2*, 96–101.

(13) Yasuda, T.; Suzuki, K.; Shimoyama, I. Automatic Transportation Of A Droplet On A Wettability Gradient Surface. In *7th lnternational Conference on Miniaturized Chemical and Blochemlcal Analysts Systems*; Squaw Valley, Callfornla, USA, 2003; pp. 1129–1132.

(14) Suda, H.; Yamada, S. Force Measurements for the Movement of a Water Drop on a Surface with a Surface Tension Gradient. *Langmuir* **2003**, *19*, 529–531.

(15) Brochard, F. Motions of Droplets on Solid Surfaces Induced by Chemical or Thermal Gradients. *Langmuir* **1989**, *5*, 432–438.

(16) Ford, M. L.; Nadim, A. Thermocapillary Migration of an Attached Drop on a Solid Surface. *Phys. Fluids* **1994**, *6*, 3183–3185.

(17) Halverson, J. D.; Maldarelli, C.; Couzis, A.; Koplik, J. A Molecular Dynamics Study of the Motion of a Nanodroplet of Pure Liquid on a Wetting Gradient. *J. Chem. Phys.* **2008**, *129*, 164708.

(18) Blake, T. D.; Coninck, J. De. The Influence of Solid-Liquid Interactions on Dynamic Wetting. *Adv. Colloid Interface Sci.* **2002**, *96*, 21–36.

(19) Blake, T. D. The Physics of Moving Wetting Lines. *J. Colloid Interface Sci.* **2006**, *299*, 1–13.

(20) Berendsen, H. J. C.; van der Spoel, D.; van Drunen, R. GROMACS: A Message-Passing Parallel Molecular Dynamics Implementation. *Comput. Phys. Commun.* **1995**, *91*, 43–56.

(21) Lindahl, E.; Hess, B. GROMACS 3 . 0 : A Package for Molecular Simulation and Trajectory Analysis. *J. Mol. Model* **2001**, 306–317.

(22) Van Der Spoel, D.; Lindahl, E.; Hess, B.; Groenhof, G.; Mark, A. E.; Berendsen, H. J. C. GROMACS: Fast, Flexible, and Free. *J. Comput. Chem.* **2005**, *26*, 1701–1718.

(23) Kong, C.-P.; Peters, E. a J. F.; Zheng, Q.-C.; de With, G.; Zhang, H.-X. The Molecular Configuration of a DOPA/ST Monolayer at the Air-Water Interface: A Molecular Dynamics Study. *Phys. Chem. Chem. Phys.* **2014**, *16*, 9634–9642.

(24) Berendsen, H. J. C.; Postma, J. P. M.; Gunsteren, W. F. van; Hermans, J. Interaction Models for Water in Relation to Protein Hydration. In *Intermolecular Forces*; The Netherlands, 1981; pp. 331–342.

(25) Mizan, T.; Savage, P.; Ziff, R. Molecular Dynamics of Supercritical Water Using a Flexible SPC Model. *J. Phys. Chem.* **1994**, 13067–13076.

(26) Martin, D. L.; Thompson, D. L.; Raff, L. M. Theoretical Studies of Termolecular Thermal Recombination of Silicon Atoms. *J. Chem. Phys.* **1986**, *84*, 4426–4428.

(27) Song, F. H.; Li, B. Q.; Liu, C. Molecular Dynamics Simulation of Nanosized Water Droplet Spreading in an Electric Field. *Langmuir* **2013**, *29*, 4266–4274.

(28) Adao, M. H.; Ruijter, M. De; Voue, M.; Coninck, J. De. Droplet Spreading on Heterogeneous Substrates Using Molecular Dynamics. *Phys. Rev. E* **1999**, *59*, 746–750.

(29) Sendner, C.; Horinek, D.; Bocquet, L.; Netz, R. R. Interfacial Water at Hydrophobic and Hydrophilic Surfaces: Slip, Viscosity, and Diffusion. *Langmuir* **2009**, *25*, 10768–10781.

(30) Hirvi, J. T.; Pakkanen, T. a. Molecular Dynamics Simulations of Water Droplets on Polymer Surfaces. *J. Chem. Phys.* **2006**, *125*, 144712.

(31) Wang, F.; Wu, H. Pinning and Depinning Mechanism of the Contact Line during Evaporation of Nano-Droplets Sessile on Textured Surfaces. *Soft Matter* **2013**, *9*, 5703–5709.

(32) Moumen, N.; Subramanian, R. S.; Mclaughlin, J. B. Experiments on the Motion of Drops on a Horizontal Solid Surface Due to a Wettability Gradient. *Langmuir* **2006**, *22*, 2682–2690.

(33) Chakraborty, M.; Ghosh, U. U.; Chakraborty, S.; Dasgupta, S. Thermally Enhanced Droplet Motion on Gradient Surfaces. *RSC Adv.* **2015**.

(34) Engin, O.; Villa, A.; Sayar, M.; Hess, B. Driving Forces for Adsorption of Amphiphilic Peptides to the Air-Water Interface. *J. Phys. Chem. B* **2010**, *114*, 11093–11101.

(35) Ruijter, M. J. De; Blake, T. D.; Coninck, J. De. Dynamic Wetting Studied by Molecular Modeling. *Langmuir* **1999**, 7836–7847.

(36) Giorgino, T. Computing 1-D Atomic Densities in Macromolecular Simulations: The Density Profile Tool for VMD. *Comput. Phys. Commun.* **2014**, *185*, 317–322.

(37) Bertrand, E.; Blake, T. D.; Coninck, J. De. Influence of Solid-Liquid Interactions on Dynamic Wetting: A Molecular Dynamics Study. *J. Phys. Condens. Matter* **2009**, *21*, 464124.

(38) Ahmadi, A.; Najjaran, H.; Holzman, J. F.; Hoorfar, M. Two-Dimensional Flow Dynamics in Digital Microfluidic Systems. *J. MICROMECHANICS MICROENGINEERING* **2009**, *19*, 1–7.

(39) Subramanian, R. S.; Moumen, N.; McLaughlin, J. B. Motion of a Drop on a Solid Surface due to a Wettability Gradient. *Langmuir* **2005**, *21*, 11844–11849.

(40) Cox, B. R. G. The Dynamics of the Spreading of Liquids on a Solid surface.Part1. Viscous FLow. *J . Fluid Mech.* **1986**, *168*, 169–194.

(41) Ahmadi, A.; Holzman, J. F.; Najjaran, H.; Hoorfar, M. Electrohydrodynamic Modeling of Microdroplet Transient Dynamics in Electrocapillary-Based Digital Microfluidic Devices. *Microfluid Nanofluid* **2011**, *10*, 1019–1032.

(42) Carlson, B. A.; Bellani, G.; Amberg, G. Measuring Contact Line Dissipation in Dynamic Wetting. *Linn´ e Flow Centre, KTH Mech. SE-100 44 Stock. Sweden* **2011**, *192*.

(43) Duvivier, D.; Blake, T. D.; Coninck, J. De. Toward a Predictive Theory of Wetting Dynamics. *Langmuir* **2013**, *29*, 10132–10140.

(44) Bertrand, E.; Blake, T. D.; Coninck, D. Influence of Solid–Liquid Interactions on Dynamic Wetting : A Molecular Dynamics Study. *J. Phys. Condens. Matter* **2009**, *21*, 14pp.

(45) Seveno, D.; Dinter, N.; Coninck, J. De. Wetting Dynamics of Drop Spreading . New Evidence for the Microscopic Validity of the Molecular-Kinetic Theory. *Langmuir* **2010**, *26*, 14642–14647.

(46) Seveno, D.; Ogonowski, G. Liquid Coating of Moving Fiber at the Nanoscale. *Langmuir* **2004**, *20*, 8385–8390.